# Ultrasensitive biosensor based on Nd:YAG waveguide laser: Tumor cell and Dextrose solution

Guanhua Li,[1] Huiyuan Li,[1] Rumei Gong,[1] Yang Tan,[2,*] Javier Rodríguez Vázquez de Aldana[3] Yuping Sun,[1] and Feng Chen[3]

[1]*Department of Respiration, Jinan Central Hospital Affiliated to Shandong University, Jinan, Shandong 250013, P.R.China*
[2]*School of Physics, State Key Laboratory of Crystal Materials, Shandong University, Jinan 250100, China*
[3]*Departamento Física Aplicada, Facultad Ciencias, Universidad de Salamanca, Salamanca 37008, Spain*
***Corresponding author:** tanyang@sdu.edu.cn*

This work demonstrates the Nd:YAG waveguide laser as an efficient platform for the bio-sensing. The waveguide was fabricated in the Nd:YAG crystal by the cooperation of the ultrafast laser writing and ion irradiation. As the laser oscillation in the Nd:YAG waveguide is ultra-sensitivity to the external environment of the waveguide. Even a weak disturbance would induce a large variation of the output power of the laser. According to this feature, the Nd:YAG waveguide coated with Graphene and $WSe_2$ layers is used as substrate for the microfluidic channel. When the microflow crosses the Nd:YAG waveguide, the laser oscillation in the waveguide is disturbed, and induces the fluctuation of the output laser. Through the analysis of the fluctuation, the concentration of the dextrose solution and the size of the tumor cell are distinguished.

## 1. INTRODUCTION

Distinguishing the individual diseased cell from the normal cells is crucial for the medical inspection and progression of the disease, especially for the tumor cells [1,2]. For this purpose, the microfluidic flow cytometry is developed for accurate detection at the single – cell level, which exhibits distinct advantages including dynamic cell manipulation, isolation, and repetitive usage [3,4]. The sensing element with the high sensitivity is the key component for microfluidic flow cytometry. Up to now, various technologies have been presented for optical flow sensing of a single cell, including flow cytometers and optofluidic circuits [5,6]. These technologies distinguish cells depending on the blocking or absorption of the detecting light, which is induced by the cell. For example, in the microfluidic channel, the cell crossing the two-dimensional material (2D materials) changes the optical absorption of 2D materials [7-9]. Via the variation of the optical absorption (or the power of the detecting light), the size of cells can be distinguished. However, the variation of the optical absorption is weak, which is hard to make an accurate detection. Therefore, there is a continuous motivation to amplify the fluctuation of the optical signal, and further improve the sensitivity of the sensing element for a single cell.

The solid-state laser is ultrasensitive to the optical loss in the resonant cavity. Even a slight variation of the optical loss will trigger a large change of the power of the output laser. It seems the solid-state laser system may be used as a potential candidate for the amplifier of the variation of the loss induced by optical absorption. Recently, the fast development of the waveguide laser makes this proposal achievable. The waveguide laser is a laser system using the waveguide as the resonant cavity and the gain medium [10,11]. The dimension of the waveguide is small (cross-sectional area less than a hundred square micron, length around several millimeters), which is suitable for the combination with the 2D materials [12,13] and the microfluidic channel [14,15]. It has been reported that the 2D materials coated on the surface of waveguide can be used as the saturable absorber for the Q-switched laser emission [12,13]. Besides, the waveguide combined with the 2D materials can also be used for the sensing of the microflow in the microfluidic channel [14-16].

In this work, we consisted a highly sensitive biosensor based on the 2D materials and waveguide laser. The Graphene and $WSe_2$ layers were placed between the microfluidic channel and the waveguide laser, which work as both the sensing medium for the microstreaming and the absorber in the waveguide laser. It is experimentally proved that the fluctuation of the optical absorption of Graphene and $WSe_2$ is amplified by the laser oscillation process in the waveguide, which provides a high sensitive sensing of the live cell and the dextrose solution. Utilizing this device, the size of the cell is ultrafast distinguished, and the concentration of the dextrose solution is detected.

## 2. EXPERIMENTS

Figure 1(a) displayed the schematic diagram of the biosensor. The waveguide was fabricated in the neodymium doped yttrium aluminum garnet (Nd:YAG) crystal by the cooperation of the ultrafast laser writing and the ion irradiation. Detailed information of the waveguide fabrication has been reported in Ref. [17]. Graphene and $WSe_2$ layer were produced by the chemical vapor deposition (CVD) and transferred onto the surface of Nd:YAG waveguide. Graphene and $WSe_2$ were stacked together consisting a G/W heterostructure, in order to achieve a higher optical absorption [18]. Figure 1(b) shows the high-resolution transmission electron microscopy (HRTEM) image of G/W heterostructure. Due to the lattice mismatching of Graphene and WSe$_2$, the lattice of heterostructure displays a Moiré pattern. A microfluidic channel with the width of 100 µm and thickness of 50 µm was coated onto the G/W heterostructure (Fig. 1(c) and (d)).

This biosensor operates in two modes, so called passive and active biosensor. 1) The passive biosensor uses a continuous laser at the wavelength of 1064 nm as the detecting light, and the information of the microflow is detected via the fluctuation of the output light. The power of the detecting light is 10 mW. Through a spherical convex lens (focus length of 25 mm), the detecting light is coupled into the waveguide. 2) The active biosensor generates the laser oscillation at 1064 nm in the Nd:YAG waveguide, and detects the microflow by the fluctuation of the waveguide laser emission at the wavelength of 1064 nm. During the experiments, a polarized light beam at a wavelength of 810 nm from a tunable CW Ti:Sapphire laser (Coherent MBR 110) is coupled into the waveguide as the pumping laser. The output laser at 1064 nm from the waveguide was collected by a long work distance microscope objective (MO, 20 ×, N.A. = 0.4).

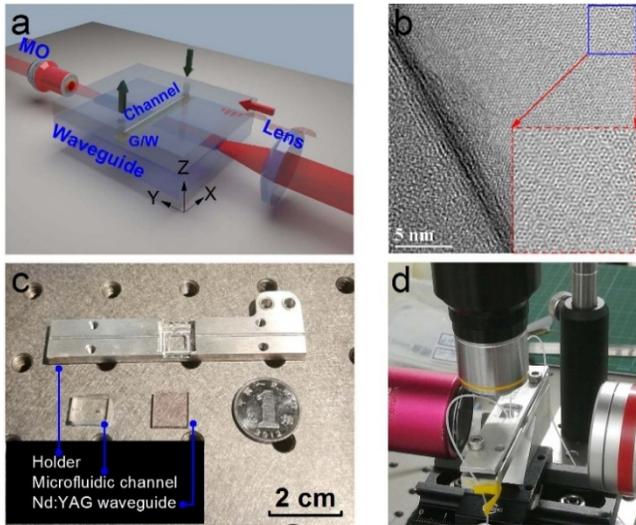

**Fig. 1.** (a) Schematic diagram of the biosensor. MO: Microscope objective. (b) HRTEM image of G/W heterostructure. (c) Photograph of the microfluidic channel, Nd:YAG waveguide and holder. (d) Photograph of the assembled biosensor.

## 3. RESULTS AND DISCUSSION

### 3.1, Passive biosensor

Figure 2(a) shows the refractive index distribution on the longitudinal section of the Nd:YAG waveguide, which has a step-like shape. The propagation mode of the guided light at the wavelength of 1064 nm was also displayed in Fig. 2(a). The intensity of the light was concentrated near the surface, and the evanescent field of the guided light has an overlapping with the G/W heterostructure. Through the interaction with heterostructure, the guided light is absorbed by the heterostructure. We detected the propagation loss of the waveguide with the different polarization of guided light at the wavelength of 1064 nm. As shown in Fig. 2(b), without the heterostructure (blue dots), the loss of the waveguide is a constant of 0.4 dB. With the heterostructure (red dots), the loss has the maximum value of 1 dB at the $s$-polarization (parallel to the waveguide surface) and the minimum value of 0.45 dB at the p-polarization (vertical to the waveguide surface). It demonstrate the G/W heterostructure has the polarization dependent absorption. In the following work, we only used the light with $s$-polarization as the detecting light.

When the microflow with different refractive index crosses the waveguide surface, the intensity of the evanescent field is changed, leading to the variation of the absorption of the heterostructure. As a result, the microflow can be detected by the fluctuation of the power of the guided light in the waveguide [9,16]. Based on this biosensor, we used a continuous laser with the power of 10 mW at 1064 nm to detect the refractive index of the air, water, and dextrose solution. As shown in Fig. 2(c), there is an obviously fluctuation of the output power with the switching of the air the liquid, corresponding to the absorption coefficient of 0.9992 dB/cm (air) and 1.1991 dB/cm (water). However, the dextrose solution with different concentration did not have an obvious change. And the absorption coefficient was slightly tuned between 1.210 dB/cm (0.6 %) and 1.257 dB/cm (5 %) in Fig. 2(d). In order to have a high resolution of the sensing, the slight difference of the absorption needs to be amplified.

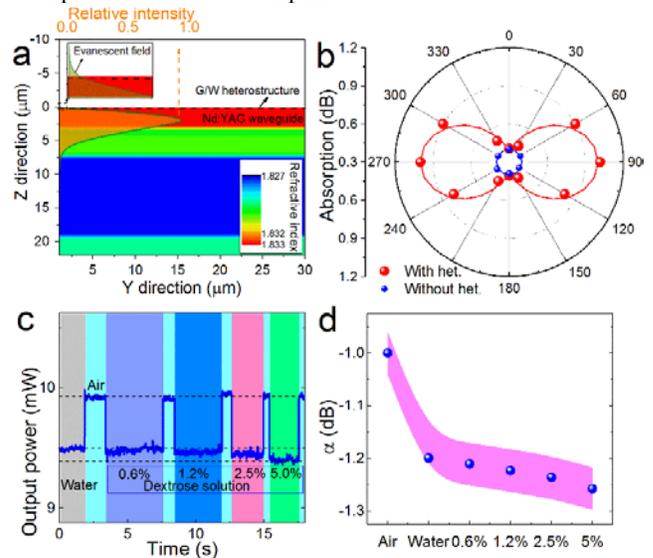

**Fig. 2.** (a) The refractive index distribution of the Nd:YAG waveguide, and the propagation mode of the guided light at the wavelength of 1064 nm. (b) The polar image of the output light power along with the polarization variation. The wavelength of the detecting light is 1064 nm. (c) Real-time signal of different concentrations of dextrose solution. (d) The absorption coefficient corresponding to air, water and dextrose solution.

### 3.2, Characteristic of the Nd:YAG waveguide laser

The laser oscillation in the Nd:YAG waveguide was excited by a continuous 810 nm laser. Figure 3(a) shows the spectrum and the near-field modal profile of the output laser at the wavelength of 1064 nm. The performance of the Nd:YAG waveguide laser is sensitive to the loss of the waveguide. To discuss the relationship of the laser performance and the loss, the loss of the Nd:YAG waveguide was artificially changed from 0.45 dB to 1.5 dB, via changing the overlay length ($L$) of the G/W heterostructure. As displayed in Fig. 3(b), there are great change of the threshold ($P_{th}$), the slope efficiency ($\Phi$) and the output power of the Nd:YAG waveguide laser, along with the variation of the loss. With the increasing of the loss, $P_{th}$ was increased from the 26 mW to 84 Mw [Fig. 3(c)], and $\Phi$ [Fig. 3(d)] was rapidly decreased from 63% to 4.5 %, leading to a dramatic change of the maximum output power [Fig. 3(e)] from 50.4 mW to 0.4 mW.

As a four level system, the relationship of the loss ($a$) and $P_{th}$ / $\Phi$ of the Nd:YAG waveguide laser can be expressed as equations below [19]:

$$P_{th} = \frac{hcA_{eff}}{2\eta\sigma_e\tau\lambda_p}\delta = C_1\delta \quad (1)$$

$$\Phi = \frac{\eta(T_1+T_2)\lambda_P}{\lambda_L}\frac{1}{\delta} = C_2\frac{1}{\delta} \quad (2)$$

$$\delta = 2\alpha L - \ln[(1-T_1)\times(1-T_2)] \quad (3)$$

where $h$ is the Planck's constant; $c$ is the light velocity in the vacuum; $\lambda_L$ and $\lambda_P$ are the wavelengths of the laser and pump beams, respectively; $\sigma_e$ is the stimulated emission cross section of Nd:YAG crystal; $\tau$ is the fluorescence lifetime; $A_{eff}$ is the effective pump area; $\eta$ is the fraction of absorbed photons that contribute to the population of the $^4F_{3/2}$ metastable state; $T_1$, $T_2$ are the transmittance of the end-faces; $\delta$ is the round-trip cavity loss.

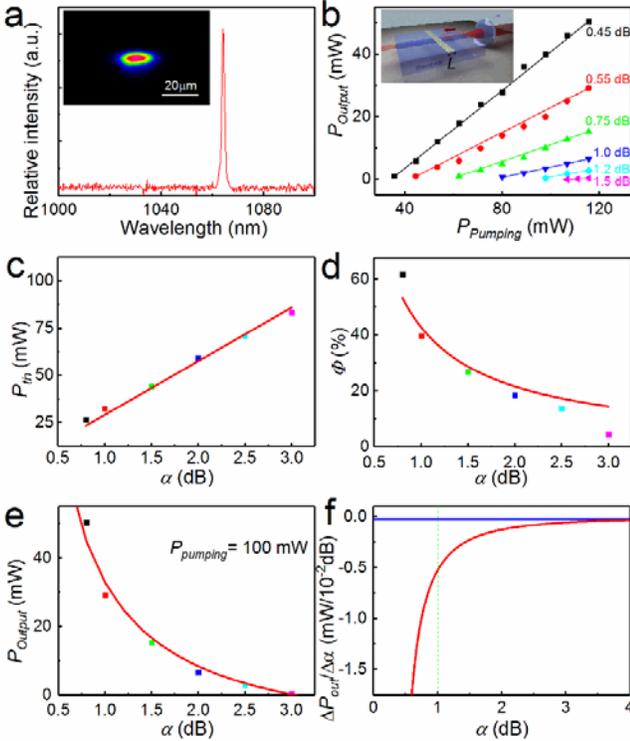

**Fig. 3.** (a) The emission spectrum of the output laser. Inset is the measured near-field modal profile of the emitted laser from the Nd:YAG waveguide. (b) The output power of the Nd:YAG waveguide as a function of the pumping power. The variations of $P_{th}$ (c), $\Phi$ (d), $P_{out}$ (e). (f) Calculated variations of $P_{out}$ per 0.01 dB in the active (red line) and passive (blue line) biosensor.

As some parameters in Eq. (1), (2) are invariable, we further simplified these equations by introducing constants of $C_1$ and $C_2$. And the value of $C_1$ and $C_2$ are obtained via fitting the measured results in Fig. 3(c) - (e). Based on Eq. (1) – (3), the maximum output power of the waveguide laser can be expressed as the following.

$$P_{out} = \Phi(P_{pump} - P_{th}) = C_2\frac{1}{\delta}(P_{pump} - C_1\delta) \quad (4)$$

$$= C_2\frac{1}{2\alpha L - \ln[(1-T_1)\times(1-T_2)]}P_{pump} - C_2C_1$$

To obtain the slewing rate of $P_{out}$, we take the partial of $P_{th}$ with respect to $a$ in Eq. (4). With the $P_{pump}$ of 100 mW (at the wavelength of 810 nm), the maximum power of the Nd:YAG waveguide laser is 50.4 mW (at the wavelength of 1064 nm), and the calculated change of $P_{out}$ per 0.01 dB (red line) is shown in Fig. 3(f). It seems the slewing rate has an exponential change along with $a$, and $P_{out}$ has a higher sensitivity at the low loss. For our biosensor ($a$ = 1 dB), the slewing rate is ~ 0.52 mW per 0.01 dB [green line in Fig. 3(f)].

With the same conditions used in the passive biosensor, the slewing rate of the output laser per 0.01 dB (blue line) is also displayed in Fig. 3(f). In this situation, the variation of the output laser is only decided by $\Delta a$. Compared with the active biosensor, the slewing rate is much lower in the passive biosensor. It demonstrates the difference of the absorption is amplified through the laser oscillation process.

### 3.3 Active biosensor

The active biosensor is used to detect the same liquid displayed in Fig. 2(c). As shown in Fig. 4(a), there is a larger difference of the output power with the different liquid. For the conversion of the water and air, the maximum variation of the power is ~ 2.47 mW, which is much higher than the passive biosensor (~ 0.45 mW). Besides, the dextrose solution with different concentration (0.6% – 5%) can be clearly distinguished. To compare the efficiency of the passive and active biosensor, we list the output power of detecting light corresponding to the refractive index of the different liquid in Fig. 4(b). The sensitivity of the active and passive biosensor is 10 mW/RIU and 1.4 mW/RIU, respectively. It demonstrates that the fluctuation of the optical intensity is amplified by the laser oscillation in the waveguide, leading to a higher sensitivity of the biosensor.

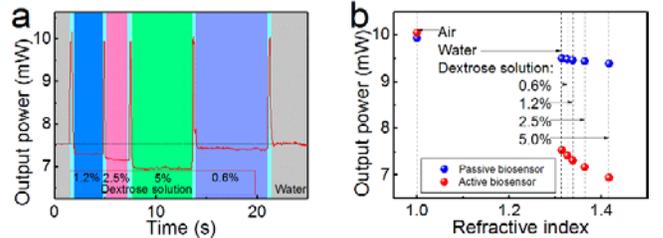

**Fig. 4.** (a) Real-time signal of different concentrations of dextrose solution in the active biosensor. (b) The output power of detecting light corresponding to the refractive index of the different liquid.

The active biosensor was used to distinguish the tumor cell and the PMMA ball. The tumor cells and PMMA balls have the diameter of ~ 20 μm [Fig. 5(a)] and ~ 10 μm, respectively. When a tumor cell or PMMA ball passes the waveguide laser through the microfluidic channel, the intensity of the evanescent field is increased. As a result, the power of the output laser has a rapid drop. We detected the power of the output laser by an ultra-fast photoelectric probe and displayed it on an oscilloscope (MSO/DPO5000B). As shown in Fig. 5(b), sharp dips were observed when an object crossing the waveguide laser. However, the intensity of dips were different, which depends on the types of objects (PMMA ball or tumor cell) [Fig. 5(b) and 5(c)]. Utilizing this feature, the tumor cell can be distinguished from the mixed solution of PMMA balls and tumor cells.

Figure 5(d) displays the Real-time signal of the mixed solution of PMMA balls and tumor cells. Obviously, there are two different dips with the distinguished intensity.

Through the counting of dips with different intensity, the concentration and type of the cell or ball in the solution could be obtained.

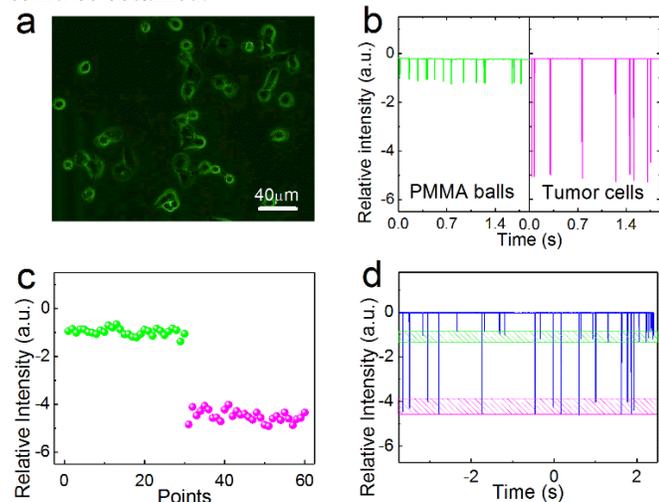

**Fig. 5.** (a) Microphotograph of the tumor cells. (b) Real-time signal of the solution of PMMA balls and tumor cells, respectively. (c) Repeatability of the active biosensor measurement. (d) Real-time signal of the mixed solution of PMMA balls and tumor cells.

## 4. CONCLUSION

We demonstrate the waveguide laser as an efficient platform for bio-sensing. The biosensor was constructed by Nd:YAG waveguide, G/W heterostructure, and a microfluidic channel. The microflow in the channel tuned the optical absorption of the G/W heterostructure, which is amplified by the laser oscillation in the Nd:YAG waveguide. The sensitivity of this biosensor is 10 mW/RIU. Based on this biosensor, we distinguished the the concentration of the dextrose solution and tumor cells. This work paves a novel way for the live cell detection on a chip.

**Funding.**
This work is supported by the National Natural Science Foundation of China (no. 11535008). Y.T. acknowledges the financial support from Young Scholars Program of Shandong University (no. 2015WLJH20).

**REFERENCES**
1. A. B. Chinen, C. M. Guan, J. R. Ferrer, J. R. Ferrer, S. N. Barnaby, T. J. Merkel, and C. A. Mirkin, "Nanoparticle probes for the detection of cancer biomarkers, cells, and tissues by fluorescence," Chem. Rev., **115**(19), 10530-10574 (2015).
2. V. Backman, M. B. Wallace, L. T. Perelman, J. T. Arendt, R. Gurjar, M. G. Müller, Q. Zhang, G. Zonios, E. Kline, T. McGillican, S. Shapshay, T. Valdez, K. Badizadegan, J. M. Crawford, M. Fitzmaurice, S. Kabani, H. S. Levin, M. Seiler, R. R. Dasari, I. Itzkan, J. Van Dam, M. S. Feld, "Detection of preinvasive cancer cells," Nature **406**, 35–36 (2000).
3. T. D. Chung, H. C. Kim, "Recent advances in miniaturized microfluidic flow cytometry for clinical use," Electrophoresis, **28**(24): 4511-4520 2007.
4. D. Huh, W. Gu1, Y. Kamotani, J. B Grotberg and S. Takayama "Microfluidics for flow cytometric analysis of cells and particles," Physiol. Meas., **26**(3): R73 (2005).
5. M. Kim, D. J. Hwang, H. Jeon, K. Hiromatsu and C. P. Grigoropoulos, "Single cell detection using a glass-based optofluidic device fabricated by femtosecond laser pulses," Lab Chip, **9**(2): 311-318 (2009).
6. Y. C. Tung, M. Zhang, C. T. Lin, K. Kurabayashi, S. J. Skerlos, "PDMS-based opto-fluidic micro flow cytometer with two-color, multi-angle fluorescence detection capability using PIN photodiodes," Sensor Actuat. B-Chem., **98**(2): 356-367 (2004).
7. P. K. Ang, A. Li, M. Jaiswal, Y. Wang, H. W. Hou, J. T. L. Thong, C. T. Lim, and K. P. Loh, "Flow sensing of single cell by graphene transistor in a microfluidic channel," Nano lett., **11**(12): 5240-5246 (2011).
8. S. Y. Yang, S. K. Hsiung, Y. C. Hung, C. M. Chang, Teh-Lu Liao and Gwo-Bin Lee, "A cell counting/sorting system incorporated with a microfabricated flow cytometer chip," Meas. Sci. Technol., **17**(7): 2001 (2006).
9. F. Xing, G. X. Meng, Q. Zhang, L. T. Pan, P. Wang, Z. B. Liu, W. S. Jiang, Y. Chen, and J. G. Tian, "Ultrasensitive flow sensing of a single cell using graphene-based optical sensors," Nano lett., **14**(6): 3563-3569 (2014).
10. C. Grivas, "Optically pumped planar waveguide lasers, Part I: Fundamentals and fabrication techniques," Prog. Quantum Electron. **35**(6), 159–239 (2011).
11. F. Chen and J. R. Vázquez de Aldana, "Optical waveguides in crystalline dielectric materials produced by femtosecond - laser micromachining," Laser Photonics Rev. **8**(2), 251–275 (2014).
12. Y. Tan, C. Cheng, S. Akhmadaliev, S. Zhou, and F. Chen, "Nd:YAG waveguide laser Q-switched by evanescent-field interaction with graphene," Opt. Express **22**(8), 9101-9106 (2014).
13. Y. Tan, S. Akhmadaliev, S. Zhou, S. Sun, and F. Chen, "Guided continuous-wave and graphene-based Q-switched lasers in carbon ion irradiated Nd:YAG ceramic channel waveguide," Opt. Express **22**(3), 3572–3577 (2014).
14. R. Osellame, V. Maselli, R. M. Vazquez, R. Ramponi, and G. Cerullo, "Integration of optical waveguides and microfluidic channels both fabricated by femtosecond laser irradiation," Appl. Phys. Lett., **90**(23): 231118 (2007).
15. V. Maselli, J. R. Grenier, S. Ho, "Femtosecond laser written optofluidic sensor: Bragg grating waveguide evanescent probing of microfluidic channel," Opt. Express, **17**(14) 11719-11729 (2009).
16. Y. Tan, R. He, C. Cheng, D. Wang, Y. Chen, and F. Chen, "Polarization-dependent optical absorption of $MoS_2$ for refractive index sensing," Sci. Rep-uk, **4** 7523 (2014).
17. J. Lv, Z. Shang, Y. Tan, J. R. V. de Aldana, and F. Chen, "Cladding-like waveguide fabricated by cooperation of ultrafast laser writing and ion irradiation: characterization and laser generation," Opt. Express, **25**(16), 19603-19608. (2017).
18. Y. Tan, X. Liu, Z. He, Y. Liu, M. Zhao, H. Zhang, and F. Chen, "Tuning of Interlayer Coupling in Large-Area Graphene/$WSe_2$ van der Waals Heterostructure via Ion Irradiation: Optical Evidences and Photonic Applications," ACS Photonics, **4**(6), 1531–1538 (2017).
19. E. Lallier, J. P. Pocholle, M. Papuchon, M. P. De Micheli, M. J. Li, Q. He, D. B. Ostrowsky, C. Grezes-Besset, and E. Pelletier, "Nd: MgO: LiNbO/sub 3/channel waveguide laser devices," IEEE J. Quantum Electron. **27**, 618 (1991).